\documentclass[a4paper,USenglish]{lipics-v2018}

\usepackage{microtype}

\usepackage{booktabs} 

\usepackage{verbatim}
\usepackage {graphicx}
\usepackage {hyperref}
\usepackage {paralist}
\usepackage{amssymb}
\usepackage{xcolor}
\usepackage{xspace}

\usepackage{algorithm}
\usepackage{algpseudocode}
\usepackage{amsmath}

\usepackage{soul}

\algrenewcommand\algorithmicprocedure{}
\algtext*{EndProcedure}
\algtext*{EndWhile}
\algtext*{EndFor}
\algtext*{EndIf}
\algblockdefx[NAME]{START}{END}%
   [3][Unknown]{#2 #1(#3)}%
   {}
\algtext*{END}

\algnewcommand\And{\textbf{ and }}
\algnewcommand\Or{\textbf{ or }}

\newcommand{\thesystem}{\textsc{Spectrum}\xspace}
\newcommand{\mtwopaxos}{$M^2P$\textsc{axos}\xspace}
\newcommand{\caesar}{\textsc{Caesar}\xspace}

\title{\thesystem: A Framework for Adapting Consensus Protocols}

\author{Balaji Arun}{Virginia Tech, Blacksburg, Virginia, United States}{balajia@vt.edu}{}{}

\author{Sebastiano Peluso}{Virginia Tech, Blacksburg, Virginia, United States}{peluso@vt.edu}{}{}

\author{Binoy Ravindran}{Virginia Tech, Blacksburg, Virginia, United States}{binoy@vt.edu}{}{}

\authorrunning{Balaji Arun; Sebastiano Peluso; Binoy Ravindran}

\Copyright{Balaji Arun; Sebastiano Peluso; and Binoy Ravindran}

\subjclass{CCS $\rightarrow$ Computer systems organization $\rightarrow$ Dependable and fault-tolerant systems and networks}


\category{}

\relatedversion{}

\supplement{}

\funding{}

\nolinenumbers 
\hideLIPIcs  

\begin{document}

\maketitle

\begin{abstract}
There exists a plethora of consensus protocols in literature. The reason is that there is no one-size-fits-all solution, since every protocol is unique and its performance is directly tied to the deployment settings and workload configurations. Some protocols are well suited for geographical scale environments, e.g., leaderless, while others provide high performance under workloads with high contention, e.g., single leader-based. Thus, existing protocols seldom adapt to changing workload conditions. 

To overcome this limitation, we propose \thesystem, a consensus framework that is able to switch consensus protocols at runtime, to enable a dynamic reaction to changes in the workload characteristics and deployment scenarios. With this framework, we provide transparent instantiation of various consensus protocols, and a completely asynchronous switching mechanism with zero downtime. We assess the effectiveness of \thesystem via an extensive experimental evaluation, which shows that \thesystem is able to limit the increase of the user perceived latency when switching among consensus protocols.
\end{abstract}

\section{Introduction}
\label{sec:intro}

Consensus is the problem of finding a common value among a set of values proposed by several entities~\cite{FLP}. Consensus underlies many distributed systems and provides fault-tolerant coordination for activities such as leader election~\cite{leader-election}, operation ordering for execution on shared and replicated objects~\cite{Burrows:2006:CLS:1298455.1298487, Bolosky:2011:PRS:1972457.1972472, elastic-paxos}, and distributed transaction processing~\cite{gray2006consensus}.


Consensus algorithms are hard to design and implement correctly. Therefore, despite the existence of many consensus algorithms in literature, most well-known distributed systems have either adopted (Multi-)Paxos~\cite{DBLP:journals/tocs/Lamport98} or Raft~\cite{raft}. 
Some users of Multi-Paxos and Raft include ~\cite{baker2011megastore, Hunt:2010:ZWC:1855840.1855851,Corbett:2012:SGG:2387880.2387905} and~\cite{cockroachdb, etcd}, respectively.




Multi-Paxos, Raft, and some other protocols~\cite{DBLP:journals/dc/Lamport06, lamport2005generalized} elect a designated node as the leader, and that node is responsible for deciding the order of all proposed messages. The limitations of this approach include scalability bottleneck under heavy load at the leader; additional latency, especially in the wide area, for clients at non-leader nodes due to communication with the leader; and loss of availability when the leader node crashes. 


To address the limitations of the single leader-based approach, a number of efforts have focused on eliminating the need for a single leader by allowing multiple nodes to operate as leaders at the same time~\cite{DBLP:conf/sosp/MoraruAK13,DBLP:conf/osdi/MaoJM08,alvin, m2paxos,caesar}. 
Most of these works exploit the characteristics of submitted commands and allow commutative commands to be ordered in different ways on different nodes. The state of the system after the execution of any two commutative commands in any order will reach the same eventual state, and hence nodes submitting commutative commands do not need to compete for ordering them~\cite{lamport2005generalized}.




However, the performance of leaderless protocols strictly depends on the contention level, defined as the amount (in percentage) of non-commutative commands submitted by the application. It has also been experimentally shown in~\cite{caesar} that these protocols could be outperformed by single leader-based ones, e.g., Multi-Paxos, when contention is very high.

Moreover, even within the lower ranges of the contention spectrum (< 30\%), different leaderless protocols have different optimum contention levels (See Figure~\ref{fig:conflict-spectrum}). Therefore, a rough classification of \emph{leaderless} and \emph{leader}-based protocols is not enough to assess their performance characteristics. 
On the other hand, system conditions such as network topology, latency between nodes, and network and node faults affect the capabilities of consensus protocols to a large extent. Works such as~\cite{marandi2010ring} address some of these deployment scenarios. 

Since there is no one-size-fits-all consensus protocol to address different workloads and deployment scenarios, we propose \thesystem, a novel, general consensus framework that provides a protocol-agnostic switching scheme that can be bootstraped with an oracle to tailor for various use cases. One use case is tolerating changing contention levels and network latencies and providing the optimum user-perceived latency with the right consensus protocol at any point in time. \thesystem can switch consensus protocols online, without any downtime, in a way completely oblivious to the clients, and while tolerating faults.

~\\
{\bf Contributions.}  The main contributions of this work are:
\begin{compactitem}
	\item[-] Classification of existing consensus protocols based on their characteristics, and motivation for \thesystem (Section~\ref{sec:motivation}). 
	\item[-] A switching mechanism that is able to coordinate among nodes to enable an oblivious transition. The subsystem, called \emph{Meta-Consensus}, is capable of reaching a non-blocking agreement among nodes on the specific switch to be performed, e.g., from Multi-Paxos to \mtwopaxos~\cite{m2paxos} (Section~\ref{sec:base}).
	\item[-] Overview of the proof of correctness of \thesystem (Section~\ref{sec:correctness}).
	\item[-] Experimental evaluation that shows the effectiveness of \thesystem (Section~\ref{sec:eval}).
\end{compactitem}

\section{Motivations for a Consensus Framework}
\label{sec:motivation}
The need for a consensus framework to support consistent and performant coordination is due to the fact that in literature there exist a plethora of consensus solutions, each of which is 
optimized for some specific configurations of workload and deployment scenarios.

Over the past decade, multiple leaderless protocols have been introduced each providing different performance guarantees for various usecases. Most of these protocols implement the Generalized Consensus specification, where the total order is obtained only among non-commutative commands. Hence, the performance of such protocols have largely been dependent on the amount of contention level, which is measured in terms of percentage of conflicts among commands from different nodes, in the system. This observation enabled grouping protocols into families by their similarities in performance. Specifically, the contention level determined, in many cases, the family of protocols that stood out from others for providing the lowest values of latency and the highest values of throughput. This clearly motivated the design of \thesystem, a framework that facilitates transition between consensus protocols to achieve the highest possible performance.

\begin{figure}
\center
\includegraphics[scale=0.6]{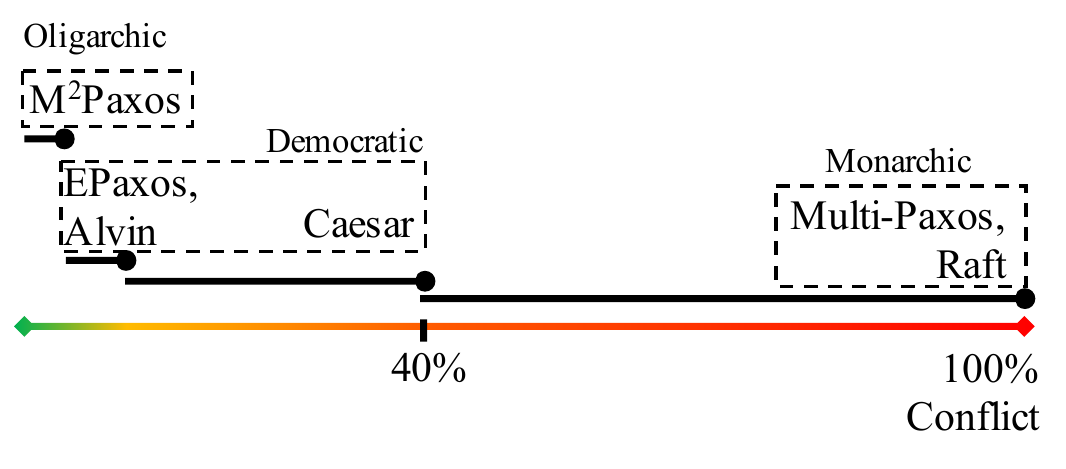}
\caption{Optimums for some consensus protocols in the conflict spectrum}
\label{fig:conflict-spectrum}
\end{figure}
~\\
{\bf Families of consensus based on contention.} Our categorization led to three \emph{consensus families}: \emph{monarchic}, \emph{oligarchic}, and \emph{democratic} (Fig.~\ref{fig:conflict-spectrum}). 

A \emph{monarchic} protocol, as the name suggests, relies on a single node for finalizing the decision of commands. 
Examples of protocols that belong to the monarchic category are (Multi-)Paxos~\cite{lamport2001paxos}, Raft~\cite{raft}, and S-Paxos~\cite{BielyMSS12}. This category also includes protocols that do not always rely on a leader for decisions, but they ultimately do so to solve conflicts that arise in case of contention, e.g., Generalized Paxos~\cite{lamport2005generalized}.

When contention is low, having a single leader limits the performance of the system, which depends on the efficiency of the leader itself.  In situation where commands mostly or always commute, a set of leader nodes can be established for autonomously deciding their own commands, given a predefined concept of ownership. The subset of leader nodes take decisions quickly and independently of other leaders, as long as no interaction is needed among the leaders.
This is the spirit of the \emph{oligarchy}, in which power rests with a small number of people. Examples of \emph{oligarchic} protocols are Mencius~\cite{DBLP:conf/osdi/MaoJM08}, and \mtwopaxos~\cite{m2paxos}. 
Oligarchy works well as long as leader nodes do not contend with each other on decisions.


In contrast, \emph{democratic} protocols are well suited when the contention is not negligible but not too high. In such protocols, a node submitting a command gathers votes from quorum of participants to determine the final order of the command with respect to dependent commands that have been concurrently submitted by other nodes.  
Therefore, every command has to undergo one or more voting phases before being decided.
Examples of protocols that belong to this category are EPaxos~\cite{DBLP:conf/sosp/MoraruAK13}, Alvin~\cite{alvin}, and \caesar~\cite{caesar}.
Among these protocols, \caesar can sustain a higher degree of contention and keep the number of voting phases to one mostly under moderate contentione, e.g., from 5\% to 30\% of conflicting commands. In low contention, EPaxos is a better choice thanks to its comparative simplicity.

~\\
{\bf Need for transparent transitions} Although adaptivity in distributed systems is a well explored problem with many solutions, the design of an adaptive consensus implementation for a combination of leaderless and leader-based protocols is an unexplored area due to the following reasons. \textit{(i)} Despite the fact that consensus is a widely investigated topic, leaderless and more scalable consensus implementations appeared in the literature recently, and they are not well established solutions yet; \textit{(ii)} Reconciling the execution of protocols that adopt fast quorums with the ones that adopt classic quorums is challenging, especially if those protocols provide different guarantees in terms of safety and liveness properties, and carry different representations of the system state, e.g., dependency graph in EPaxos, ownership maps in \mtwopaxos, partial timestamp-based orders in Caesar, total order in Multi-Paxos.  

A desirable adaptive consensus should switch from one specific consensus protocol to another one at runtime, independently of whether the protocols belong to the same family or not, in order to adapt the consensus layer to the characteristic of the workload and deployment scenarios. In doing so, it is crucial that the switching mechanism provides the same safety guarantees of the protocols that it manages and coordinates, while being transparent to the user.
Moreover, the decision to switch from one protocol to another must be fault-tolerant; a decision to switch to a new protocol instance should persist even in the presence of crash faults. This means ruling out any solution where a single designated node takes the decision about the switch unilaterally.

Note that this work focuses on reconfiguration among different consensus protocols, and not within a single protocol. For instance, this work does not provide a description of how to reconfigure the internal characteristics of Caesar (leaderless) to act like a single-leader one. Instead, \thesystem composes protocols as they are, and to do so, we leverage certain ideas from existing work on reconfiguration~\cite{lamport2010reconfiguring} and builds a solution using it.  
Moreover, leaderless protocols are newer, and reconfiguration among leaderless and leader-based protocols have not been explored to the best of our knowledge. This paper aims to address this gap. Moreover, we emphasize that leaderless protocols have their advantages and advocate their use in practice, but when they fail, \thesystem can alleviate the damage. 

\thesystem provides a protocol-agnostic switching scheme that can be bootstraped with an oracle to tailor for various usecases; we present two: maximizing performance during contention and changing latency.



\section{System Model}
\label{sec:system-model}
We assume a set of nodes $\Pi = \{p_1, p_2, \ldots, p_{N}\}$ that communicate through message passing and do not have access to either a shared memory or a global clock. Nodes may fail by crashing but do not behave maliciously, i.e., they are not byzantine. A node that does not crash is called correct; otherwise, it is faulty.  We assume an asynchronous model where messages may experience arbitrarily long and unknown, although finite, delays. Furthermore, any node can send messages to and receive messages from any other node, with the following guatantees: if $p_i$ sends a message to $p_j$ and both of them are correct, then $p_j$ eventually receives that message; if $p_j$ receives a message, then it receives that message once, and that message has been sent by some process. 

Following the FLP~\cite{FLP} result, we assume that the system can be enhanced with the weakest type of unreliable failure detector~\cite{DBLP:journals/tcs/GuerraouiS01} that is necessary to implement a leader election service~\cite{luisBook}. This is a reasonable assumption to guarantee that eventually there are no two correct processes that trust different correct processes as the leader. This has also been assumed by the authors of all the consensus protocols that we can instantiate in \thesystem, e.g., ~\cite{caesar},~\cite{m2paxos},~\cite{DBLP:conf/sosp/MoraruAK13}, to guarantee that eventually there will be one leader for deciding the order of a command. 

In addition, we assume that at least a strict majority of nodes, i.e., $\left\lfloor\frac{N}{2}\right\rfloor+1$, is correct; otherwise, it is impossible to solve consensus given the assumptions on the unreliability of the failure detector and the asynchrony of the network~\cite{FLP}. We name \textit{classic quorum} ($\mathcal{CQ}$), or more simply \textit{quorum}, any subset of $\Pi$ with size at least equal to $\left\lfloor\frac{N}{2}\right\rfloor+1$.

It should be noted that quorums that are as small as classic quorums are not sufficient to allow protocols like EPaxos, \caesar, or Generalized Paxos to take fast decisions in two communication steps. Therefore, we assume that at least a \textit{fast quorum} ($\mathcal{FQ}$) of $\left\lceil\frac{3N}{4}\right\rceil$ nodes is correct, to extract the best performance from all protocols that are instantiated in \thesystem; however, this assumption is not required to guarantee correctness.

We follow the definition of Generalized Consensus~\cite{lamport2005generalized}: each node can propose a command, e.g., $\begingroup\color{red}\blacktriangle\endgroup$, via the \textsc{Propose}$(\begingroup\color{red}\blacktriangle\endgroup)$ interface, and nodes decide command structures $C$-$struct$ $cs$ via the \textsc{Decide}$(cs)$ interface. The specification is such that: commands that are included in decided $C$-$structs$ must have been proposed (\textit{Non-triviality}); if a node decided a $C$-$struct$ $cs$ at any time, then at all later times it can only decide $cs \bullet \sigma$, where $\sigma$ is a sequence of commands (\textit{Stability}); if $\begingroup\color{red}\blacktriangle\endgroup$ has been proposed then $\begingroup\color{red}\blacktriangle\endgroup$ will be eventually decided in some $C$-$struct$ (\textit{Liveness}); and two $C$-$structs$ decided by two different nodes are prefixes of the same $C$-$struct$ (\textit{Consistency}). The operator $\bullet$ stands for \emph{append}, as defined in~\cite{lamport2005generalized}.

For simplicity, we use the notation \textsc{Decide}$(\begingroup\color{red}\blacktriangle\endgroup)$ for the decision of a command $\begingroup\color{red}\blacktriangle\endgroup$ on a node $p_i$, with the following semantics: a sequence of consecutive calls of \textsc{Decide}, i.e.,~\textsc{Decide}$(\begingroup\color{red}\blacktriangle\endgroup)$, \textsc{Decide}$(\begingroup\color{blue}\blacksquare\endgroup)$, $\cdots$ , \textsc{Decide}$(\begingroup\color{green}\blacklozenge\endgroup)$ on $p_i$ is equivalent to the call of \textsc{Decide}$(\begingroup\color{red}\blacktriangle\endgroup\bullet \begingroup\color{blue}\blacksquare\endgroup \bullet \cdots \bullet \begingroup\color{green}\blacklozenge\endgroup)$.

We say that two commands $\begingroup\color{red}\blacktriangle\endgroup$ and $\begingroup\color{blue}\blacksquare\endgroup$ are \textit{non-commutative}, or \textit{conflicting}, and we write $\begingroup\color{red}\blacktriangle\endgroup\sim\begingroup\color{blue}\blacksquare\endgroup$, if the results of the execution of both $\begingroup\color{red}\blacktriangle\endgroup$ and $\begingroup\color{blue}\blacksquare\endgroup$ depend on whether $\begingroup\color{red}\blacktriangle\endgroup$ has been executed before or after $\begingroup\color{blue}\blacksquare\endgroup$. It is worth noting that, as specified in~\cite{lamport2005generalized}, two $C$-$structs$ are still the same if they only differ by a permutation of non-conflicting commands.

\section{\thesystem}
\label{sec:base}

\thesystem provides dynamic and transparent switches, fault-tolerant and non-blocking decisions, which enables adaptivity to different workload and deployment configurations, by adopting a novel switching solution that is based on consensus. In particular, the core idea is abstracting the whole switching and adaptation functionalities into a higher level consensus layer. Such a layer coordinates the switch between consensus protocols, while minimizing the unavoidable increase in user-perceived latency during transitions. Having a meta layer of consensus coordinating a switch from one consensus protocol to another, and taking care of the lifetime of the protocols, is a novel challenge.
\begin{figure}
\center
\begin{subfigure}{0.4\textwidth}
	\includegraphics[scale=.8]{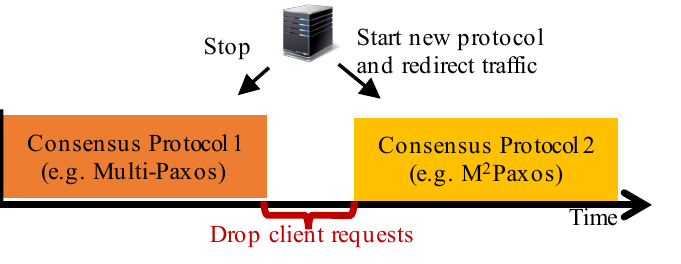}    
	\caption{naive \emph{stop-and-restart}}
    \label{fig:naive-protocol}
\end{subfigure}
\begin{subfigure}{0.4\textwidth}
    \includegraphics[scale=.8]{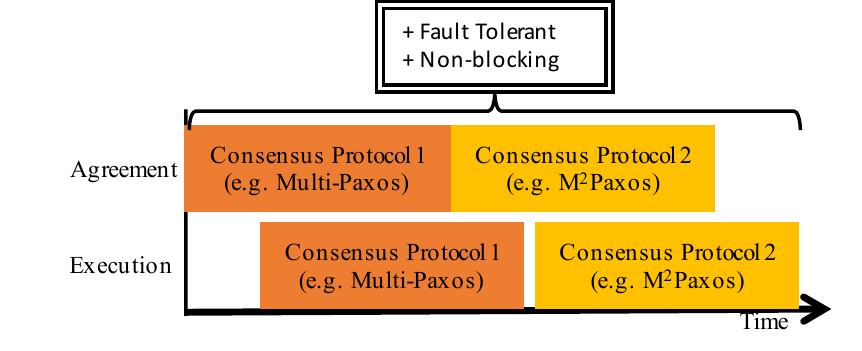}
    \caption{\thesystem approach}
    \label{fig:spectrum-overview}
\end{subfigure}
\caption{\emph{Stop-and-restart} vs \thesystem. The former lacks fault-tolerance, non-blocking behavior and oblivious transition, while the latter provides these capabilities.}
\label{fig:naive-vs-spectrum}
\end{figure}
~\\
{\bf Intuition.} Before understanding the main intuition behind \thesystem, let us consider a simple but intuitive switching scheme where we have some node triggering and coordinating a transition. The coordinator node stops the execution of one consensus protocol and starts the execution of a new protocol, like in Figure~\ref{fig:naive-protocol}. We call this \emph{stop-and-restart} solution. There are two main problems in such a setup: firstly, the coordination of the transition is not fault-tolerant, leaving the system vulnerable to crashes during the switching of a protocol; secondly, it is not clear whether commands that are submitted during the transition have to be rejected or not and, if not, which is the protocol in charge of deciding those commands. On the other hand, if commands are rejected, the system would suffers from a downtime. 


To solve this, we leverage a result presented in~\cite{cheap-recovery}, where the authors prove that in state machine replication, the separation of agreement and execution is a necessary and sufficient condition to enable lazy recovery. Lazy recovery is a flexible solution to solve the problem of on-demand instantiation of replica nodes; the aim is to activate the minimal number of replicas first, and as they fail, activate the backup replicas. 

Note that implementing state machine replication is equivalent to implementing consensus~\cite{lamport2005generalized}. Furthermore, the problem of on-demand instantiation in state machine replication is similar to the problem of instantiation of a new consensus protocol and the transfer of the state of execution.
This leads to our solution, whose high level idea is depicted in Figure~\ref{fig:execution}.

Solving consensus involves finding an agreement on the order of commands, and then executing them --- which may simply mean delivering commands to the upper layer in the software stack ---  in that order. 
By recognizing that fault-tolerant transition among consensus protocols can actually be guaranteed via consensus itself, \thesystem adopts a solution where a meta consensus layer triggers and coordinates a switch, and uses the separation between agreement and execution components in both the pre-switch and post-switch consensus instances, in order to guarantee continuous delivery of commands. The correctness of such a modularized approach has been also formally proved by the work in~\cite{speculative-linearizability}.

The benefits of \thesystem are remarkable. Commands are never rejected during the transition period since the agreement layer provides continuity. However, they can experience an increase in latency due to a possible gap in time between the end of a pre-switch execution and the beginning of a post-switch execution. This is unavoidable since commands have to be delivered according to an order such that commands processed by the pre-switch agreement should be ordered before commands processed by the post-switch agreement.

The rest of this section describes the \thesystem and its subsystems in detail with examples. The algorithmic pseudocode is deferred to Appendix~\ref{sec:detailed-algorithm}, due to space constraints. \thesystem is a switching framework that requires an oracle that triggers the switch; such an oracle is presented in Appendix~\ref{sec:oracle}.

\subsection{Meta-Consensus}
\label{sec:metaconsensus}
With the expectation of ensuring full transparency to the user, it is crucial that \thesystem provides the same guarantees of the underlying consensus protocols. Therefore, an implementation of consensus itself as the core mechanism is the best fit for coordinating the protocol transitions in a non-blocking, fault-tolerant, and consistent way. This would also enable the modularization of the consensus protocols themselves, which can be integrated into the higher level consensus-based coordination module, which we call \emph{Meta-Consensus}, as plugins.
 

\emph{Meta-Consensus} module comprises of two components: the \emph{agreement component} and the \emph{management component}. The \emph{agreement component} is the core consensus that decides on the next \texttt{Switch} command, which contains information on the switch itself (e.g., next consensus protocol), to be applied in a fault-tolerant manner. The \emph{management component} is responsible for executing the decisions from the agreement component. Particularly, it manages the instances of the consensus protocols that are involved in the switch; it coordinates those instances in such a way that the commands that are submitted by the clients are correctly ordered, even if they have been submitted during the transition.

\begin{figure}
\center
\includegraphics[scale=.5]{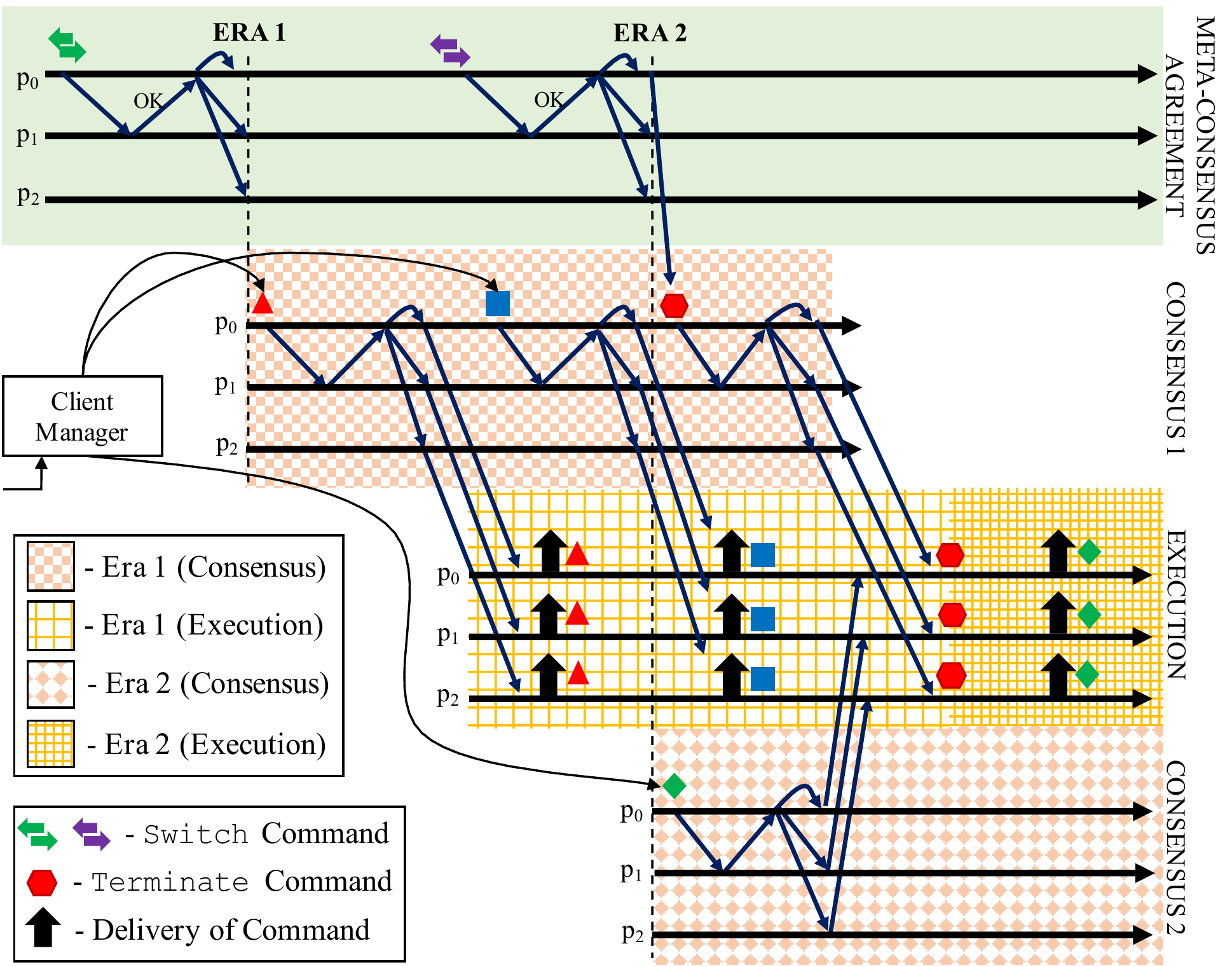}
\caption{An example execution of \thesystem.}
\label{fig:execution}
\end{figure}

An example run of \thesystem is depicted in Figure~\ref{fig:execution}. It shows all the steps that the system takes for commands that are submitted by clients before, during, and after a transition: the commands $\begingroup\color{red}\blacktriangle\endgroup$, $\begingroup\color{blue}\blacksquare\endgroup$, and $\begingroup\color{green}\blacklozenge\endgroup$, respectively. It also shows the execution of a transition itself. As described in the intuition part of Section~\ref{sec:base}, we divide the run of a consensus protocol in two parts, agreement and execution.

The figure depicts three nodes that participates in the agreement and execution of commands, and it is organized as follows. The first stripe  represents the steps of the \emph{agreement component} of \emph{Meta-Consensus}; the second and the fourth ones represent the agreement parts of the instance of consensus before the switch, i.e., pre-switch consensus, and the instance of consensus after the switch, i.e., post-switch consensus, respectively; the third one represents the execution part of the pre-switch consensus followed by the execution part of the post-switch consensus.

As a note, \emph{management component} is not explicitly shown in the figure, although its steps are properly discussed through the description in the next Section~\ref{sec:management}.

\subsubsection{Agreement Component}
\label{sec:agreement}
Most of the existing consensus protocols in literature --- both single leader-based and leaderless --- are viable candidates as implementations of the \emph{agreement component}. Indeed, they have to implement agreement among nodes to decide the \texttt{Switch} command. The \texttt{Switch} command is a \thesystem command that encodes information necessary to perform the switch. It contains the identifier for the next consensus protocol that the \emph{management component} must instantiate and redirect client traffic to. The result of execution of this command informs whether the switch was successful or not.


The \emph{agreement component} implements a leader-based consensus algorithm, due to its simplicity, for deciding on the \texttt{Switch} command. 
Node $p_0$ is the leader that triggers the transition in Figure~\ref{fig:execution}. It asks all the nodes to accept the \texttt{Switch} command, and when it receives a quorum of acknowledgement messages, the decision becomes safe and is not lost even under faults, given the assumptions in Section~\ref{sec:system-model}. The leader, then, informs all the nodes that the transition has to happen. In the figure, \thesystem takes steps for the \texttt{Switch} command twice. The first one, is just to show that the procedure of the switch can be used to initialize the first instance of a consensus protocol at the beginning of a run, while the second one is to show how to trigger an actual transition.

The agreement and the execution of a command that has been submitted before the execution of a transition proceeds normally according to the active consensus instance. For example, command $\begingroup\color{red}\blacktriangle\endgroup$ in the figure, undergoes an agreement phase first (second stripe) and an execution phase then (third stripe). For simplicity in the representation, we adopt monarchic protocols, i.e., Multi-Paxos, as both pre-switch and post-switch consensus protocols. Therefore, the agreement of command $\begingroup\color{red}\blacktriangle\endgroup$, as well as all the commands that are submitted by the clients, follows the algorithm of Multi-Paxos in the figure.


Moreover, it is sufficient to use a simpler protocols such as Multi-Paxos or Raft for the \emph{agreement component}. \emph{Meta-Consensus} agreement only receives a command when there is a need to switch and is not susceptible to concurrent requests from nodes.
 



\subsubsection{Management Component}
\label{sec:management}
Once the \emph{agreement component} decides on the next \texttt{Switch} command, it sends the command to the \emph{management component} for execution. The execution of \texttt{Switch} involves the creation of a new consensus protocol instance, as specified in the command, and redirection of client requests to the newly created instance. 
The \emph{management component} also serves as the entry point for client commands into the client-facing consensus instances that it manages. Note that client commands do not pass through the \emph{agreement component} of \emph{Meta-Consensus}, but directly enter the management component. Only \texttt{Switch} commands pass through the agreement component.

When the \emph{management component} receives the \texttt{Switch} from the \emph{agreement component}, it creates the instance of the new consensus protocol as defined in the command, and initializes it. Thanks to the properties of consensus, the \texttt{Switch} is delivered to all the correct instances of the \emph{management component} in the system, hence all the correct nodes creates and initializes the instance of the new consensus protocol that \thesystem is going to switch to.
In Figure~\ref{fig:execution}, when $p_0$ informs all that the switch as to happen, nodes independently create a new instance of consensus in the fourth stripe, which can start processing command $\begingroup\color{green}\blacklozenge\endgroup$.


Through the rest of the paper, we call the period during which a consensus protocol instance lives as an \emph{era}. 
An \emph{era} is started whenever a \texttt{Switch} is executed and a new instance of a protocol is created as a result, and hence it is strictly tied to a consensus instance. 
For example, from left to right in Figure~\ref{fig:execution}, the first vertical line denotes the start of \emph{era} $1$, and the second one denotes the start of \emph{era} $2$.
The core idea is that whenever a new \emph{era} is started in a node, then all other nodes in the system will eventually start the new \emph{era}, thanks to the properties of consensus that characterize the \emph{agreement component}. Those nodes are the participants in the post-switch consensus instance.

In addition to the aforementioned tasks, the \emph{management component} is also responsible for ensuring the in-order execution of the client commands with respect to different client-facing consensus protocols. Recall that a new \emph{era} is created whenever a \texttt{Switch} is executed, and consequently, after that the ordering of new commands happens in the new \emph{era}.
This means that, at any given time during the switch, both the consensus instances are ordering commands that are submitted from the clients. The instance in the old \emph{era} orders commands that were pending at the time the transition started, while the instance in the new \emph{era} orders commands that are submitted from the start of the transition on. This way, the management component enables continuity of the service, with no downtime in ordering the commands.
For example, in Figure~\ref{fig:execution}, command $\begingroup\color{blue}\blacksquare\endgroup$, which was already in the system when the switch has started, is decided in \emph{era} $1$, while command $\begingroup\color{green}\blacklozenge\endgroup$, which had been submitted by the client after the switch,
is decided in \emph{era} $2$.

However, the execution happens only in one \emph{era} at a time to establish a total order between commands in different \emph{eras}. To facilitate this, the commands in the new \emph{era} that are ready for execution are buffered until the previous \emph{era} terminates. This means that command $\begingroup\color{green}\blacklozenge\endgroup$ in the figure, which belongs to \emph{era} $2$, cannot be executed until commands $\begingroup\color{red}\blacktriangle\endgroup$ and $\begingroup\color{blue}\blacksquare\endgroup$, which belong to \emph{era} $1$, have been. 
The \emph{management component} takes care of that, and ensures that the commands are dispatched for execution first in the order that is decided by \emph{Meta-Consensus} and then, within an \emph{era}, by the order that is decided by the respective instance of consensus that is running in that \emph{era}.
The way \thesystem defines the completion of an \emph{era} depends on the characteristics of the consensus protocols that are involved in the switch, and it is described in Section~\ref{sec:switch}.

\subsection{Switching between Consensus Categories} 
\label{sec:switch}
\emph{Meta-Consensus} manages the instantiation of a new \emph{era} and the switch of the traffic to that \emph{era}. However, as mentioned in the previous section, the \emph{management component} of \emph{Meta-Consensus} has to ensure that commands in a new \emph{era} are executed only after commands in any previous \emph{era}.
Therefore, the core problem of the switching mechanism is the identification of completion of a previous \emph{era}; that is, the \emph{management component} should know when commands in a previous \emph{era} have been executed.
To enable this functionality, we adopt a technique used in Stoppable Paxos~\cite{stoppable-paxos}. The idea is to propose a special \texttt{Terminate} command to the previous \emph{era} protocol that acts as a marker indicating the end of commands in that \emph{era}. Once \texttt{Terminate} command is delivered to the management component, it can start executing commands in the new \emph{era}. 

Despite the differences in the nature of existing families of consensus (see Section~\ref{sec:motivation}), \thesystem proposes a uniform method for handling the delivery of \texttt{Terminate} commands. For clarity, we describe two mechanisms -- one for transition from single leader-based and another from leaderless protocol -- and later combine them into one, for simplicity.



\subsubsection{Transition from Single Leader-based Consensus}
\label{sec:single-to-leaderless}
In \emph{monarchic} protocols, such as Multi-Paxos and Raft, there exists only one node, i.e. the leader, that decides on the proposed commands on behalf of all the nodes in the system. Meanwhile, other nodes in the system just forward their commands from the clients to the leader and acknowledge proposals from the leader. Therefore, it is sufficient to send the \texttt{Terminate} command to a leader node, and let that node establish the termination of an instance in an \emph{era}. In Figure~\ref{fig:execution}, $p_0$ first requests a quorum of nodes to accept the termination of the consensus instance at \emph{era} $1$, and then forces them to terminate that instance. Meanwhile, new client commands are ordered in the new \emph{era}, but are buffered for execution until the old \emph{era} terminates.


When the \texttt{Terminate} command is delivered to the \emph{management component} at a node, it starts the transition to execute commands in the new \emph{era}, e.g., command $\begingroup\color{green}\blacklozenge\endgroup$ in the figure.
This is because the \texttt{Terminate} command ensures the total order of the commands between two consecutive \emph{eras}. The properties of consensus ensure that commands  $\begingroup\color{red}\blacktriangle\endgroup$ and $\begingroup\color{blue}\blacksquare\endgroup$ belonging to Consensus 1 in figure have already been processed and executed by a node at the time the final decision for \texttt{Terminate} is processed by that node. 


It should be noted that the decision on the termination is fault-tolerant even though only one node takes care of the decision. This is because that the node is anyway the leader of a consensus instance and, as such, it must ensure that the decision is stable before forcing it to all the others.


\subsubsection{Transition from Leaderless Consensus}
\label{sec:leaderless-to-single}

Leaderless (e.g., \emph{democratic}, \emph{oligarchic}) protocols typically implement the Generalized Consensus specification, in that the total order is obtained only among non-commutative commands. In addition, every node can propose and decide on the outcome of their commands. This entails that the \texttt{Terminate} command must find a total order among all the commands submitted to the system by all nodes. 

In general, the commands submitted to Generalized Consensus protocols must embed information defining their commutativity. In practice, this is achieved by using the identities of the application objects that the commands operate on. Thus, by including the identities of all application objects, the \texttt{Terminate} command can establish a total order among all the commands submitted to the system. One optimization is to include the range of identities rather than every single identity in the command, which practical systems support~\cite{cockroachdb}.

In protocols such as \caesar and EPaxos, this will cause a \emph{dependency} on all the prior commands in the system, and hence will be delivered after all the previous commands have been delivered. On the other hand, in \mtwopaxos, the \texttt{Terminate} command will prompt one node to acquire ownership of all the objects, thus establishing a total order. Note that this method is applicable for any Generalized Consensus leaderless protocols.

Among leaderless protocols, one exception is Mencius~\cite{DBLP:conf/osdi/MaoJM08}, as it does not implement Generalized Consensus. Instead, it totally orders all the commands submitted to the system. Hence, the method described in previous section will instead apply for such protocols.

Note that, in practice, the \texttt{Terminate} command need not be devised specifically for a particular transition type (i.e. from leader-based or leaderless). Instead, it can include the commutativity information always and leave it to the underlying consensus protocol to use the information if required.

\section{Correctness Arguments}
\label{sec:correctness}
We show how \thesystem guarantees the properties of consensus, as described in Section~\ref{sec:system-model}. Due to space constraints we do not provide a complete formal proof. However, we provide the reader with a very intuitive explanation on the correctness of \thesystem, which partially relies on the correctness of the underlying consensus protocols.

\thesystem guarantees \textit{Non-triviality} by construction of the algorithm, which decides at least the commands that are proposed, and by the assumption on the network layer, which neither creates nor duplicates messages. \textit{Stability} is trivially guaranteed since commands are decided one-by-one, and the decisions are never retreated. Furthermore, a proposed command will be eventually decided for the following reasons: \thesystem proposes a command to one of the underlying consensus protocol instances that it is coordinating, and therefore it provides the same liveness guarantees of those protocols.

However, there are two exceptions. -- \textit{i)} the command has been proposed to an instance of consensus that might not guarantee liveness due to the amount of contention, e.g., \mtwopaxos, and hence is never learnt -- \textit{ii)} the command is learnt by an instance of consensus after the instance decides a \texttt{Terminate} command, which means that the command cannot be decided, at least in the era associated with that instance. In both cases, the command remains pending, and the client that submitted it never receives a reply for it.

The solution is relying on commands re-transmission: a client that does not receive a reply for a command in a predefined amount of time, it can resubmits the command, which is eventually decided (if had not been decided before) by an underlying consensus instance after a possible switch. It is worth noticing that \thesystem eventually performs a switch in case a liveness problem arises due to contention.

Liveness is guaranteed in case of crashes as well. The critical scenario is when the leader that coordinates a switch crashes while it is broadcasting a \textsc{ChangeEra} message to trigger a \texttt{Switch}. If that happens, then there exists at least one correct node that knows the information about the switch, since the leader has received a quorum of \textsc{AckAccept} messages that are tagged as \emph{ACK} before sending \textsc{ChangeEra}. Furthermore, no more than $min\left\{\mathcal{CQ}\right\}-1$ nodes can crashes, where $min\left\{\mathcal{CQ}\right\}$ is the minimum size of a quorum. 

\thesystem guarantees \textit{Consistency} for commands that are not submitted during a transition because the underlying consensus protocol instances guarantee  \textit{Consistency} as well. On the other hand, \thesystem guarantees that, if a command is learnt before a \texttt{Terminate} command on a node in \emph{era} $x$, then all correct nodes learn that command before \texttt{Terminate} in $x$, thanks to the \textit{Consistency} of the active consensus instance in $x$. Furthermore, if a command has been learnt after \texttt{Terminate} had been decided in \emph{era} $x$, then the command is either not decided or it has been learnt in \emph{era} $y>x$.

Since commands that are learnt by a node in \emph{era} $y$ cannot be decided by the \emph{execution} instance of $y$ until the commands that are lerant by that node in every \emph{era} $x<y$ have been decided by the execution instance of $x$, \textit{Consistency} follows.


\section{Experimental Evaluation}
\label{sec:eval}

We prototyped \thesystem in Java and incorporated the following consensus protocols into the framework: Multi-Paxos, \caesar, and \mtwopaxos. 
Each of these protocols are fundamentally diverse, and they effectively demonstrate the capabilities of \thesystem in carrying out the switch. 
Moreover, these three protocols cover the majority of the conflict spectrum (Figure~\ref{fig:conflict-spectrum}) and allow us demonstrate an important usecase for \thesystem: coping with workloads with varying contention level over time. 
To demonstrate the effectiveness of our approach, we designed three different experiments, each highlighting a unique aspect of the framework. The specifics of the experiments are discussed in the following subsections. 

For each experiment, we deployed our prototype(s) in five AWS regions around the world using the EC2 service. The regions are located as follows: two in USA (Virginia and Ohio), two in Europe (Ireland and Frankfurt), and one in Asia (Mumbai). In each site, we used a 8-core virtual machine with 32 GB of memory running Ubuntu 16.04. We colocated a client process with 50 worker threads in each virtual machine to inject requests into the prototype(s). The experiments, measure latency at different sites by injecting requests in a closed loop. This is important because each site incurs different latency to other sites, and the latency per communication step varies per site.

\subsection{Minimizing latency under conflicts}

\begin{figure*}[t]
\center
\begin{subfigure}{1\textwidth}
    \includegraphics[scale=.5]{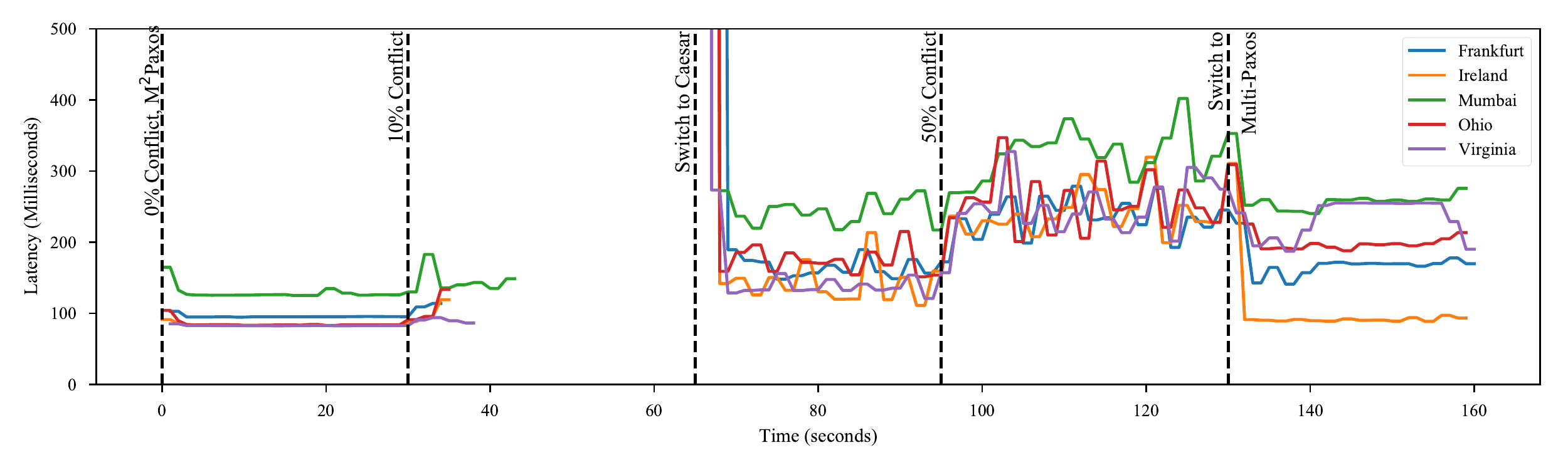}
    \caption{Scenario showing the effectiveness of the switch as the contention in the workload increases over time.}
    \label{fig:increasing-conflict}
\end{subfigure}
\begin{subfigure}{1\textwidth}
    \includegraphics[scale=.5]{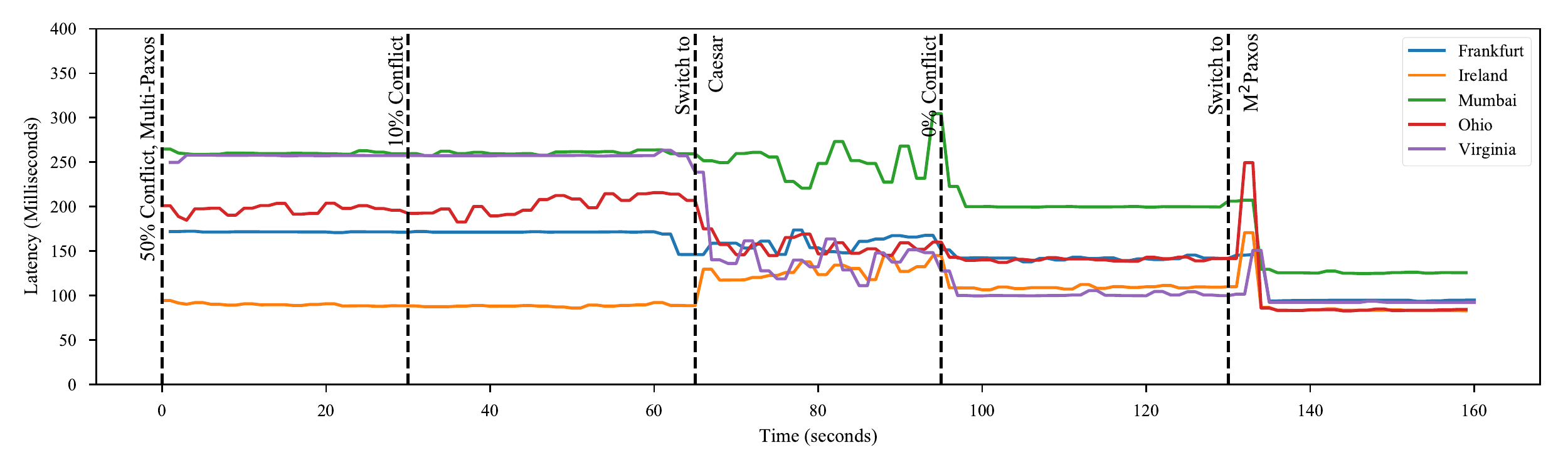}
    \caption{Scenario showing the effectiveness of the switch as the contention in the workload decreases over time.}
    \label{fig:decreasing-conflict}
\end{subfigure}
\caption{Effectiveness of \thesystem in providing minimum possible latency for any conflicting workload}
\label{fig:changing-conflict}
\end{figure*}

In this experiment, we show that \thesystem's ability to switch consensus protocols at runtime enables to provide the minimum possible user-perceived latency for any contentious workload. In other words, as the contention level in the workload changes, \thesystem can switch to a different protocol to reduce the user-perceived latency. Contention level is adjusted by changing the amount of concurrent conflicting commands entering the system. For example, at 10\% contention, every client in the system issues a non-commuting command once every ten commands (at random) that conflicts with another such concurrent command.
To show this, we devised a static oracle that observes the amount of conflict in the workload injected to the system and triggers the switch to a protocol that can better handle that particular workload. The condition for the oracle to trigger the switch depends on the percentage of conflicts: if that is in the range $[0,10)$, the oracle chooses \mtwopaxos; if in the range $[10,50)$ the oracle chooses \caesar; otherwise the oracle chooses Multi-Paxos. 




We conducted two sub-experiments using the oracle. In the first experiment, we increase the contention level over time in order to provoke a switch. The result is presented in Figure~\ref{fig:increasing-conflict}. The x-axis presents the time in seconds since the beginning of the experiment, and the y-axis shows user-perceived latency in milliseconds.

We initialize \thesystem with \mtwopaxos as the starting consensus protocol and inject a non-conflicting workload. In this case, the oracle does not trigger any switch as \mtwopaxos is the best protocol for this workload. At time $t=30s$, the amount of conflicts in the workload is increased to 10\%. At this point, \mtwopaxos experiences a livelock caused by conflicting ownership acquisitions, and thus the client requests timeout. The oracle steps in at $t=65s$ and switches to \caesar, and in few seconds, \thesystem responds and delivers commands to the client. At $t=95s$, we increase the conflict to 50\%, and \caesar starts performing poorly, but not as worse as \mtwopaxos at 10\% conflict. The oracle triggers the switch to Multi-Paxos at $t=130s$, and this reduces the latency as at this amount of conflict, a monarchic protocol is better than any other group of protocols.

In the second experiment, we reversed the process; that is, we reduced the conflicting percentage from 50\% to 0\% over a time period. The result is in Figure~\ref{fig:decreasing-conflict}. Here, the system is initialized with Multi-Paxos, and a 50\% conflicting workload is injected initially. The oracle does not trigger any switch as conditions are suitable for Multi-Paxos. At $t=30s$, the conflict percentage drops from 50\% to 10\%, and at this condition, \caesar is preferable. Therefore, at $t=65s$, the oracle triggers a switch to \caesar, and the latency drops down immediately. At $t=95s$, we reduce the conflict to 0\%, and this workload can be better served by \mtwopaxos. Therefore, such a switch is triggered at $t=130s$ by the oracle and latency drops again.

Note that we configured our oracle to perform the switch after 35 seconds of observing a new workload and keep the conflict rate same for about 60 seconds. This is done to explicitly contrast the change in latency due to the new workload before and after the switch. 

\subsection{Comparison against an alternate switching scheme}

\begin{figure}
\begin{subfigure}{0.48\textwidth}
    \includegraphics[scale=.55]{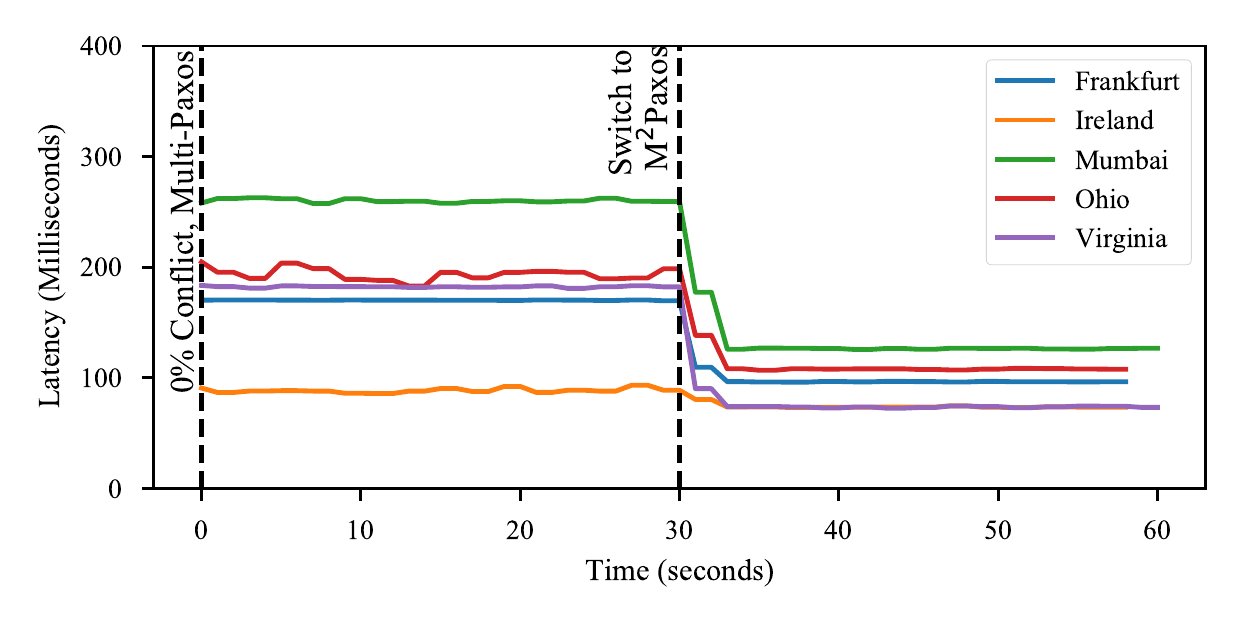}
    \caption{\thesystem}
    \label{fig:spectrum}
\end{subfigure}
\begin{subfigure}{0.45\textwidth}
    \includegraphics[scale=.55]{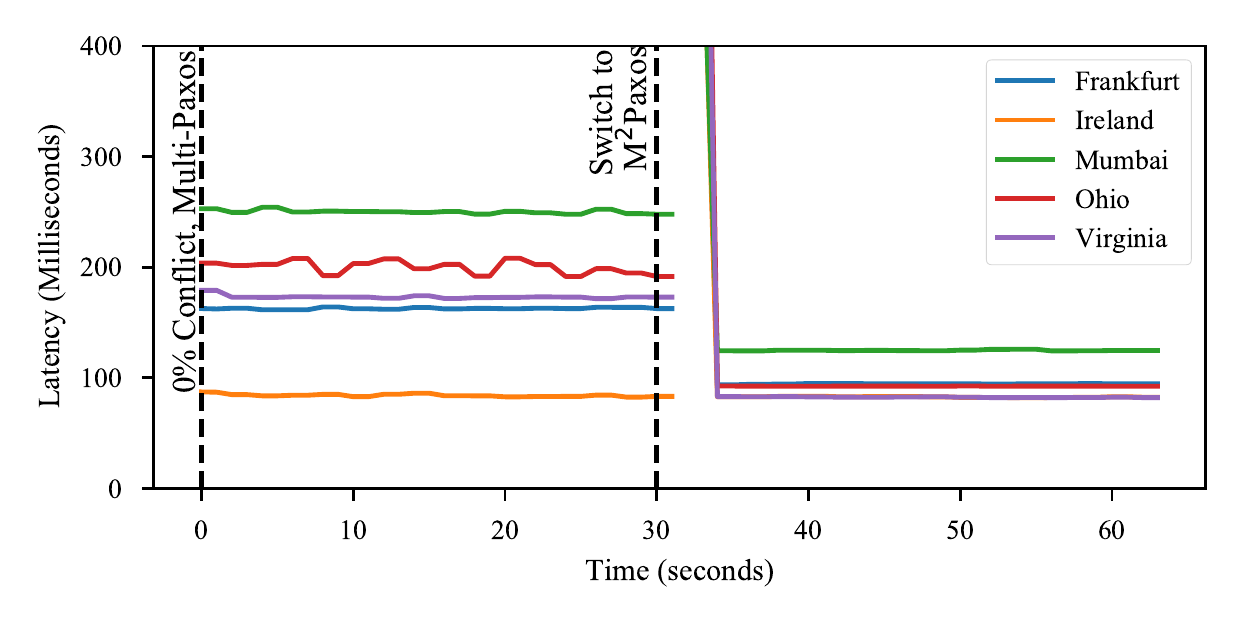}
    \caption{\emph{Stop-and-restart}}
    \label{fig:stopstart}
\end{subfigure}
\caption{\thesystem vs \emph{Stop-and-restart}. The former performs the switch oblivious to the client, while the latter drops client requests during the switch.}
\label{fig:comparison}
\end{figure}

We implemented \emph{Stop-and-restart}, an alternate switching scheme that follows the one shown in Figure~\ref{fig:naive-protocol}, to prove the value added of our solution, during a transition. In \emph{Stop-and-restart}, an external coordinator node implements the switch by first stopping the current instance of the consensus protocol, ensuring that all pending commands have been decided and executed, and then starting the new instance of a different consensus protocol. Note that, in contrast to \thesystem, \emph{Stop-and-restart} rejects any new command that is submitted to the system during the transition, and is not fault-tolerant.

In Figure~\ref{fig:comparison}, we contrast the performance of \thesystem and \emph{Stop-and-restart} during the switch from Multi-Paxos to \mtwopaxos. The workload has been kept constant with 0\% conflict throughout the experiment to show only the cost of the \texttt{Switch}. It can be observed that the switch in \thesystem is transparent to the client (Figure~\ref{fig:spectrum}). On the other hand in \emph{Stop-and-restart}, the presence of an external coordinator to handle the switch and the fact that the transition is not streamlined, causes the client requests to timeout, and once the switch completes, client requests receive their responses.

This is shown in Figure~\ref{fig:stopstart} with a period of no latency reporting (indicating the transition period) followed by a sharp spike (indicating switch completion and client response). 


%

\subsection{Fault tolerance}

We devised a specific experiment to show the fault-tolerance of \thesystem. We instrumented \emph{Meta-Consensus} to force a crash of the leader during the agreement on the switch. Specifically, right before broadcasting \textsc{Decide}. As a result, the failure detector of a correct node performs the recovery to finalize the decision of the pending \texttt{Switch} command.

\begin{figure}
\center
\includegraphics[scale=.6]{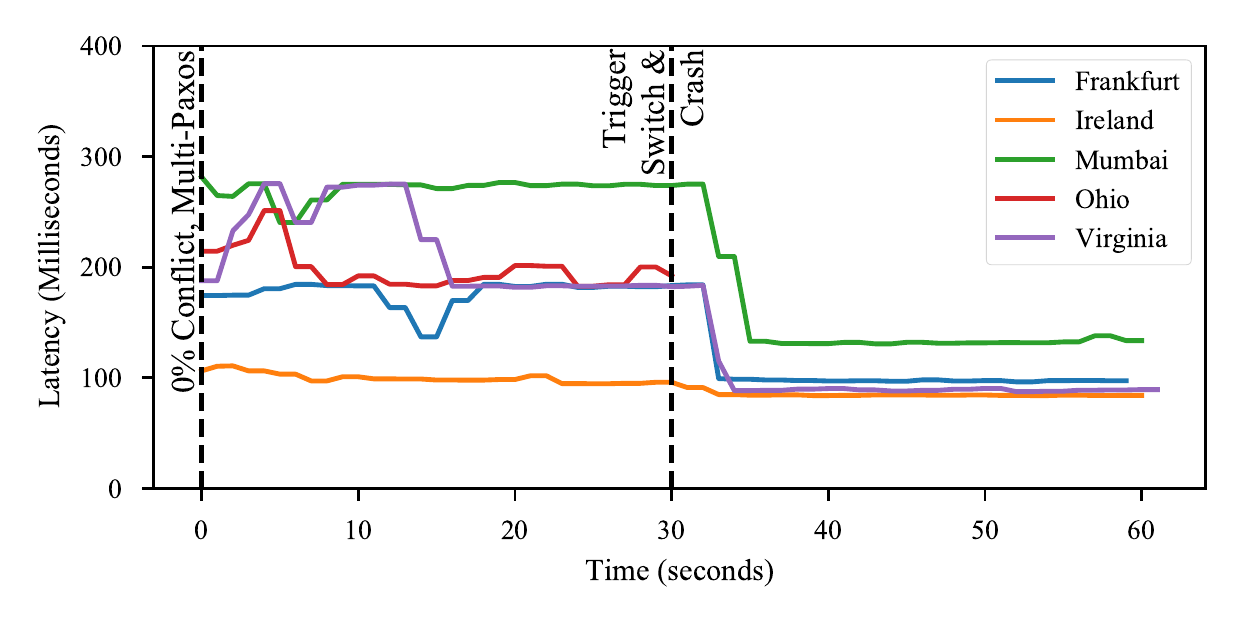}
\caption{Fault Tolerance of \thesystem.}
\label{fig:recovery}
\end{figure}

The result is shown in Figure~\ref{fig:recovery}. At $t=30$, the leader in Ohio is forcefully crashed. Due to the crash, only the clients of that node timeout, while clients of other nodes keep receiving responses from the active consensus protocol. Other clients do not observe any difference in latency since the only node that benefitted from Ohio was Virginia, e.g., having Ohio is its quorums. However, Virginia can access Frankfurt with almost the same latency as Ireland, which already was in the quorums.

\section{Related Work}
\label{sec:related-work}

Composition of consensus protocols has been investigated in~\cite{guerraoui2010next}. \texttt{Abstract} is an abstraction for designing and reconfiguring generalized state machines, by leveraging the idea of composing instances of different fault-tolerant consensus protocols. The idea is to build simpler consensus protocols each tolerating particular  system conditions such as fault models and contention, and compose them together to achieve a robust system. The downside of the \texttt{Abstract} approach is that it requires the candidate protocols to implement specific interfaces. Specifically, the candidates must be able to export (import, respectively) internal state outside (into, respectively) the protocol. This means that existing as well as new protocols must be rethought to accommodate to the abstraction.

\thesystem, on the other hand, provides isolation and composition without requiring any changes to candidate consensus protocols. Specifically, the protocols under consideration maintain state differently and thus transfer of such state from one protocol to another is a tedious, perhaps an impossible task. \thesystem treats candidate consensus protocols as black-boxes and builds a general solution such that no state transfer is necessary. This makes our framework more appealing for existing as well as new consensus designs.



In~\cite{lamport2010reconfiguring}, the authors describe algorithms for reconfiguring state machines, including the one presented in~\cite{stoppable-paxos}. The authors propose using a special \texttt{Terminate} command whose delivery will stop the active protocol, after which the reconfiguration such as number of nodes and failure tolerance can be adjusted. This is similar to the \emph{stop-and-restart} solution, with which we compared our approach. Moreover, as mentioned in Section~\ref{sec:switch}, 
\thesystem also uses the idea of using the \texttt{Terminate} command from~\cite{stoppable-paxos}. However, \thesystem  stands out distinctly due to its ability to perform the switch seamlessly while ensuring fault-tolerance. In addition, we present a complete detailed algorithm that has been implemented and experimentally evaluated, unlike~\cite{lamport2010reconfiguring, stoppable-paxos}.




{\bf Adaptivity in distributed systems.} Our solution is related to self-tunable solutions that adapt their internal mechanisms to react to changes in the environment. \thesystem specifically focuses on the mechanisms offering an effective and cheap way for the adaptation than on the oracles that trigger the adaptation. The design of the oracles is an orthogonal problem, and solutions for that can be used as plugins to the framework.

TAS~\cite{tas} is an approach for automating the elastic scaling of replicated in-memory transactional systems. \thesystem can benefit from its performance predictor that relies on the combined usage of analytical modeling and machine learning, since it is able to forecast the effects of data contention. For the same reason, the machine learning-based model of MorphR~\cite{morph-r} can be exploited by \thesystem, which finds the optimal transactional replication protocols according to the conflicts in the system. MorphR is able to choose between blocking, i.e., 2-PC, and non-blocking, i.e., total order, protocols, but it does not focus on the optimal switching mechanisms among non-blocking protocols, e.g., different consensus protocols.


\section{Conclusion}
\label{sec:conclusion}
This paper reasserts that there exist many implementations of consensus, and each one is highly optimized to outperform the others in certain specific configurations of the workload and deployment scenarios. Some provide high scalability under no contention, others do not scale well but guarantee good performance even in case of high contention. Therefore, we present \thesystem, a consensus framework that is able to switch consensus protocols at runtime, to enable a dynamic reaction to changes in system conditions, and  guarantee zero downtime during the transitions. For completeness, an oracle in presented in the Appendix.



\bibliography{references}

\appendix

\section{Detailed Algorithm}
\label{sec:detailed-algorithm}

This section presents the detailed description of \thesystem along with the algorithmic pseudocode.


The flexibility of our \thesystem approach is that any consensus protocol that exhibits certain common characteristics is a good candidate to be pluggable into the adaptive consensus system. One of the characteristics is the ability to separate execution from agreement, because \thesystem enables parallel agreement but sequential execution between protocols. The separation is required to activate the agreement of commands independently of their execution, by ensuring service continuity of the ordering mechanism even during a transition, and defer the activation of the execution only after the previous \emph{era} has terminated.

As depicted in Algorithm~\ref{alg:interface}, a consensus protocol has to implement two interfaces: \textsc{Agreement} and \textsc{Execution} interfaces. The former includes an input function to propose a new command, i.e.,~\textsc{Propose}, and an output function that is called by the consensus to notify that the order of a command has been learnt by the local node, i.e.,~\textsc{Learn}. The latter includes two input functions: one to add a new learnt command to the pool of possibly executable commands of the execution layer, i.e.,~\textsc{AppendForExecution}; another one to retrieve and delete from the pool of the execution layer the next executable command, i.e., \textsc{GetNextDeliverable}.

On the other hand, \thesystem implements the interface of consensus, with the semantics that we described in Section~\ref{sec:system-model}: \textsc{Propose} to propose a new command, and \textsc{Decide} to notify a client that the command is ready to be executed.


\begin{algorithm}[h]
\begin{algorithmic}[1]
\caption{\emph{Consensus and Execution Component Interface}}
\label{alg:interface}
{\scriptsize
\START[\textsc{Agreement}]{{\bf interface}}{}
\State {\bf in function} {\textsc{Propose}}(\emph{Cmd cmd})
\State {\bf out} {\textsc{Learn}}(\emph{Cmd cmd})
\END
\Statex
\START[\textsc{Execution}]{{\bf interface}}{}
\State {\bf in function} {\textsc{AppendForExecution}}(\emph{Cmd cmd})
\State {\bf function} {\textsc{GetNextDeliverable}} {\bf returns} \emph{Cmd cmd}
\END
}
\end{algorithmic}
\end{algorithm}

\begin{algorithm}[t]
\caption{ \emph{Era Switch: Core Protocol} (node $p_i$).}
\label{alg:coordination}
\begin{algorithmic}[1]
{\scriptsize
\START[\textsc{EraPropose}]{{\bf upon}}{\texttt{Switch} \emph{cmd}} initiated by leader $\mathcal{L}$
\State $era \gets LastDecided + 1$
\State {\bf return} \textsc{EraAcceptPhase}$(cmd, era, Epoch)$
\END
\Statex
\START[\textsc{EraAcceptPhase}]{{\bf function} \emph{Bool}}{\texttt{Switch} \emph{cmd, era, epoch}}
\State {\bf send} \textsc{EraAccept}$(\langle cmd,era,epoch \rangle)$ \textbf{to all} $p_k \in \Pi$
\State $Set$ $rep \gets$ {\bf receive}  \textsc{AckAccept}$(\langle era,epoch,cmd, -\rangle)$ {\bf from} $Quorum$
\If{$\exists \langle era,epoch, cmd, NACK\rangle \in rep$}
	\State \Return $\bot$
\Else
	\State {\bf send} \textsc{Decide}$(\langle cmd,era,epoch\rangle)$ \textbf{to all} $p_k \in \Pi$
	\State \Return $\top$
\EndIf	
\END
\State
\START[\textsc{EraAccept}]{{\bf upon}}{$\langle \texttt{Switch}$ $cmd,$ $era,$ $epoch\rangle$} {\bf from} $p_j$
\If{$Rnd[era] \leq epoch$}
	\State $Vdec[era] \gets cmd $
	\State $Rdec[era] \gets epoch$
	\State $Rnd[era] \gets epoch$
	\State {\bf send} \textsc{AckAccept}$(\langle era,epoch,cmd,ACK\rangle)$ \textbf{to all} $p_j$
\Else
	\State {\bf send} \textsc{AckAccept}$(\langle era,epoch,cmd,NACK\rangle)$ \textbf{to} $p_j$
\EndIf
\END 
\Statex
\START[\textsc{Decide}]{{\bf upon}}{$\langle  \texttt{Switch}$ $cmd,$ $era,$ $epoch\rangle$} {\bf from} $p_j$
\If{$Decided[era] = NULL$}
	\State $Decided[era] \gets cmd$
\EndIf
\END
\Statex 
\START[]{{\bf upon}}{$\exists cmd: \exists era:$ $Decided[era] = cmd$ $\land$ $era = LastDecided + 1$}
	\State {\bf trigger} \textsc{ChangeEra}$(era, cmd)$
	\State $LastDecided++$
\END
\Statex
\START[\textsc{ChangeEra}]{{\bf upon}}{era, \texttt{Switch} \emph{cmd}}
	\State {\bf send} $ActiveAgreement$.\textsc{Propose}[$\texttt{TERMINATE}$]
	\State $\langle id \rangle \gets cmd$
	\State $Consensus$ $\gets$ {\bf trigger} $\textsc{InitConsensus}[era, id]$
	\State $Executor$ $\gets$ {\bf trigger} $\textsc{InitExecutor}[era, id]$
	\State $Protocols[era] \gets \langle Consensus, Executor \rangle$
	\State $ActiveAgreement \gets Consensus$
	\If{$ActiveExecutor = null$}
		\State $ActiveExecutor = Executor$
	\EndIf
\END
}
\end{algorithmic}
\end{algorithm}

The algorithm to start and coordinate the switch of a protocol, and start a new era at process $p_i$ is presented in Algorithm~\ref{alg:coordination}. Each process maintains the following global variables: -- \textit{ActiveAgreement}, which stores the instance of the agreement implementation of the current consensus protocol; -- \textit{ActiveExecution}, which stores the instance of the execution implementation of the current consensus protocol; -- \textit{ExecId}, which stores the \emph{era} number that \thesystem is currently executing commands in; -- \textit{Protocols}, which is an array that maps \emph{eras} to a pairs of consensus protocol instances and their respective executor; -- \textit{LastDecided} is the most recent era that \thesystem has switched to; --~\textit{Epoch}, which stores the highest epoch number that $p_i$ has ever seen; -- \textit{Decided}, which is an array that maps \emph{eras} to decided switch commands.


As outlined in Section~\ref{sec:agreement}, the current leader that is in charge of executing a switch can propose a new switch by broadcasting an \textsc{EraAccept} (line 5) with the tag \texttt{Switch} and information about the next switch: the new $era$ (line 2) and the current epoch number, as well as protocol specific info that is stored in \textit{cmd}. Then it waits for a quorum of replies and, if it does not receive any reply that is tagged as \textit{NACK}, it can broadcast a \textsc{Decide} message containing the same info as before (lines 6--11).

A node accepts an \textsc{EraAccept} message, and hence replies with an \textsc{AckAccept} message that is tagged as \textit{ACK} (line 18), if the value of \textit{Rnd[era]} is not greater than the received epoch number (lines 13--14). \textit{Rnd[era]} stores the newest epoch number that is associated with \textit{era}, in which $p_i$ has participated in the run of the coordination protocol with that era. If the message has been accepted, the node updates \textit{Rnd[era]} (line 17), and it stores in \textit{Vdec[era]} the \texttt{Switch} command \textit{cmd} that it accepted for \textit{era} (line 15), and in \textit{Rdec[era]} the information that is has accepted it in that \textit{epoch} (line 16).

If a node cannot accept an \textsc{EraAccept}, it can reply with an \textsc{AckAccept} that is tagged as \textit{NACK} (lines 19--20).

When a process receives a \textsc{Decide} message for a switch command cmd and an \textit{era}, it can set the variable \textit{Decided[era]} to \textit{cmd} if that was \textit{NULL} (lines 21--23). After that, a node can change an era whenever there is a valid switch command in \textit{Decided[era]}, and \textit{era} is right the era following the current one, i.e., \textit{LastDecided} (lines 24--26).

Changing an era means proposing the \texttt{Terminate} command to the agreement implementation of the current instance of consensus (line 28), initializing new agreement and execution implementations of the next consensus instance that \thesystem has to switch to (lines 30--31), storing this information in \textit{Protocols[era]} for the new era (line 32), and update the $ActiveAgreement$ (line 33).

\begin{algorithm}
\caption{ \emph{Era Switch: Recovery} (node $p_i$).}
\label{alg:recovery}
\begin{algorithmic}[1]
{\scriptsize
\START[\textsc{EraRecovery}]{{\bf upon}}{\texttt{Switch} \emph{cmd}} initiated when leader $\mathcal{L}$ is suspected
\State $era \gets LastDecided + 1$
\State $epoch \gets ++Epoch$

\State {\bf send} \textsc{Prepare}$(\langle era, epoch\rangle)$ \textbf{to all} $p_k \in \Pi$
\State $Set$ $rep \gets$ {\bf receive}  \textsc{AckPrepare}$(\langle era, epoch, -, -\rangle)$ {\bf from} $Quorum$

\If{$\exists \langle instance, epoch, NACK, -\rangle \in replies$}
	\State {\bf return}
\Else
	\State $k$ $\gets max(\{r:\langle in,-,r\rangle \in dec \land \langle -,-,-,dec\rangle \in rep\})$
	\State $Cmd$ $toForce \gets v : \langle in,v,k\rangle \in dec \land \langle -,-,-,dec\rangle \in rep$
	\If{$toForce = null$}
		\State $Bool$ $r \gets$ \textsc{EraAcceptPhase}$(cmd, era, epoch)$
	\Else
		\State $Bool$ $r \gets$ \textsc{EraAcceptPhase}$(toForce, era, epoch)$
		\State {\bf trigger} \textsc{EraPropose}$(cmd)$
	\EndIf
	\If{$r = \bot$}
		\State {\bf trigger} \textsc{EraRecovery}$(cmd)$
	\EndIf
\EndIf	
\END
\Statex
\START[\textsc{Prepare}]{{\bf upon}}{$\langle$ $era,$ $epoch\rangle$} {\bf from} $p_j$ 
\If{$Rnd[era] < epoch$}
	\State $Rnd[era] \gets epoch$
	\State $decided \gets \{ \langle era, Vdec[era],Rdec[era] \rangle$
	\State {\bf send} \textsc{AckPrepare}$(\langle era,epoch,ACK,decided\rangle)$ \textbf{to} $p_j$
\Else
	\State {\bf send} \textsc{AckPrepare}$(\langle era,epoch,NACK,-\rangle)$ \textbf{to} $p_j$
\EndIf
\END
}
\end{algorithmic}
\end{algorithm}

When the leader is suspected to have crashed, a node in the system will try to recover any pending \texttt{Switch} command in an $era$ that was not decided by the previous leader, using the recovery procedure in Algorithm~\ref{alg:recovery}. The node recovering the \textit{cmd} increments the node's epoch \textit{Epoch} to a value higher than its current value and broadcasts a \textsc{Prepare} message with the new \textit{Epoch} and the pending \textit{era} (lines 2--3). When other nodes receive this \textsc{Prepare} message, they will respond positively if the $epoch$ is higher than what they had previously observed and include in the reply any command $cmd_j$ that they accepted for \textit{era} (lines 19--22). Once the recoverer node receives a quorum of \textsc{PrepareOK} message tagged with \textit{ACK}, it analyzes the received replies (lines 9--10). If any of the replies contain information that another node has already accepted a $cmd_j$ at an \textit{era}, then recoverer node must force it by triggering \textsc{EraAcceptPhase} for \emph{Epoch} and \textit{era}; and then, recover its command $cmd$ in another era (lines 13--15). If there is no such $cmd_j$, then the recoverer can trigger \textsc{EraAcceptPhase} for its $cmd$ (lines 11--12).

Algorithm~\ref{alg:execution}, shows the algorithm to propose and executes commands from the clients. Whenever a client proposes a command, \thesystem proposes the command to the agreement implementation of the current active consensus instance (lines 1--2). Whenever the order of a command has been decided by the agreement implementation of some consensus protocol instance (that can be one of the instances that are still running during the transition), \thesystem calls \textsc{Learn} for that command to that specific agreement implementation. This means that the command can be added to the pool of commands that might be executable for the execution implementation associated with that agreement (lines 3--5).

If there is a command in the $ActiveExecutor$ that is ready to be decided, \thesystem can call \textsc{Decide} for that command (line 11). If that command is a \texttt{Terminate}, the node points the $ActiveExecutor$ to the next available executor that was instantiated as a result of the switch (lines 7--9).

\begin{algorithm}
\caption{\emph{Universal Propose Interface} (node $p_i$).}
\label{alg:execution}
\begin{algorithmic}[1]
{\scriptsize
\START[\textsc{Propose}]{}{\emph{Cmd c}}
	\State {\bf send} $ActiveAgreement$.\textsc{Propose}[\emph{c}]
\END
\Statex
\START[\textsc{Learn}]{{\bf upon receive}}{\emph{Cmd cmd}} from protocol instance $instance$
\State $\langle -, Executor \rangle \gets Protocols[instance]$ 
\State $Executor.\textsc{AppendForExecution}[cmd]$
\END
\Statex
\START[]{{\bf upon}}{$\exists c :  c = ActiveExecutor.\textsc{GetNextDeliverable()}$ $\land$ $c$ {\bf not} $null$}
\If {$c = cmd_T$}
\State {$ExecId \gets ExecId + 1$}
\State {$ActiveExecutor \gets Procotols[ExecId].Executor$}
\Else
\State {\bf send} {$\textsc{Decide}[c]$}
\EndIf
\END
}
\end{algorithmic}
\end{algorithm}




\section{Oracle}
\label{sec:oracle}

Designing an efficient oracle is orthogonal to the problem of designing an efficient mechanism for consensus switching, and it requires further investigation that is out of the scope of this paper. However, there exists a very rich literature on self-tuning distributed systems, with a particular focus on replicated transactional systems, e.g.,~\cite{diego-partial-replication}, and distributed atomic commitment protocols, e.g.,~\cite{morph-r}. Exploring such solutions would be the first step towards an as much accurate as possible oracle that is suitable for consensus.





\begin{algorithm}[h]
\caption{Oracle Algorithm}
\label{alg:oracle}
\begin{algorithmic}[1]
{\scriptsize
\State \textsc{$QRTT_i$}: Latency to collect a simple majority quorum for $node_i$
\State \textsc{$FQRTT_i$}: Latency to collect a fast quorum for $node_i$
\State \textsc{$FRTT_i$}: Latency to forward command to leader and receive reply for $node_i$
\State \textsc{$Latency_i$}: p90 Latency at $node_i$ over a moving window of $x$ seconds
\State \textsc{$Contention$}: Average Contention over a moving window of $x$ seconds
\Statex
\START[]{{\bf ChooseBestByLatencyFor}}{\emph{Node $i$}}
\State {\bf recompute} $QRTT_i, FQRTT_i, FRTT_i$
\If {$Latency_i < FQRTT_i$} {\bf return} $\langle M^2Paxos \rangle$
\ElsIf {$Latency_i < QRTT_i + FRTT_i$} {\bf return} $\langle Caesar \rangle$
\ElsIf {$Latency_i > QRTT_i + FRTT_i $} {\bf return} $\langle Multi-Paxos \rangle$
\EndIf
\END
\Statex
\START[]{{\bf ChooseBestByContentionFor}}{\emph{Node $i$}}
\If {$Contention < 10\%$} {\bf return} $\langle M^2Paxos \rangle$
\ElsIf {$Contention < 50\%$} {\bf return} $\langle Caesar \rangle$
\ElsIf {$Contention > 50\%$} {\bf return} $\langle Multi-Paxos \rangle$
\EndIf
\END
\Statex
\START[]{{\bf upon change}}{$Latency_i : \forall i \in N \lor Contention$}
\If {$Contention$ {\bf changed}}
\State $v = Mode(\{\forall i \in N : ChooseBestByContentionFor(i)\})$
\State {\bf send} \textsc{EraPropose}[$\langle v \rangle$] to Meta-Consensus
\Else
\State $v = Mode(\{\forall i \in N : ChooseBestByLatencyFor(i)\})$
\State {\bf send} \textsc{EraPropose}[$\langle v \rangle$] to Meta-Consensus
\EndIf
\END
%
}
\end{algorithmic}
\end{algorithm}

In this section, we present a simple but capable algorithm for a switching oracle that triggers the \emph{Era-Switch} to instantiate a consensus protocol that is best suited for the current network conditions and contention level (Algorithm~\ref{alg:oracle}). Such an oracle can be plugged in into the Meta-Consensus layer (of a node) so that the oracle can monitor system conditions and take decisions accordingly.

Our oracle design requires few input parameters that can be provided by 
the Meta-Consensus layer. The parameters are listed in lines 1--5 of the algorithm. The two high-level parameters the oracle takes into account are the average perceived command latencies at each node and the average contention level in the whole system. Since different consensus protocols have different requirement on quorums, we need to take into account specific quorum latencies when deciding the optimal protocol for a particular latency. Thus, the oracle requires three latency parameters: $QRTT_i$ is applicable for $M^2$Paxos; $FQRTT_i$ is used by Caesar; and $QRTT_i$ and $FRTT_i$ is used by Multi-Paxos. Note that these latencies can by computed at each node using the round-trip latencies to all other nodes, and thus requires no extra steps. For example, $QRTT_i$ is the maximum of the round-trip latencies to all the quorum nodes for node $i$. To account for network perturbations, it is advisable to record the $90^{th}$ percentile ($p90$) latencies. $Contention$ represents the average contention in the entire system.

The oracle works as follows. A change in the $p90$ latencies or the average contention over a time period (say 10 seconds) witnessed by the nodes (line 5), triggers the search for the optimal algorithm. If both the parameters change, contention is prioritized. \emph{ChooseBestByContentionFor} makes picks the right protocol for the current contention levels. This constants used in this function have been derived based on experience and works well in most cases. 

The contention based decision maker falls short when latency increases despite unchanged contention. Consider an example: even though the $Contention$ is less than $50\%$, where Caesar would work well, the average latency could be well beyond $QRTT + FRTT$ due to a partitioned node since Caesar requires more nodes in a fast quorum than Multi-Paxos. In this case, Multi-Paxos can actually provide a lower latency than Caesar. The\emph{ChooseBestByLatencyFor} function is called per node $i$ to collect the best algorithm at each node, and the mode of the results is used as the new algorithm. The \emph{ChooseBestByLatencyFor} compares the command latencies and various the quorum latencies (lines 1--3), to find the optimal algorithm that reduces the overall system latency.

Moreover, it is important for the change functions to be triggered an adequate duration apart (say 20 seconds), in order for the system to adapt. In addition, when accommodating other protocols, only the above two functions require modification, and the change is trivial.

\end{document}